\documentclass[showpacs,amsmath,amssymb,twocolumn,aps]{revtex4}
\usepackage{graphicx}% Include figure files
\usepackage{epsfig}
\usepackage{pstricks}
\usepackage{psfrag}
\usepackage{dcolumn}% Align table columns on decimal point
\usepackage{bm}% bold math

 \def\comment#1{}

\begin{document}
\title{Defect induced melting of vortices in high-${\bf T_c} $
    superconductors: A model based on continuum elasticity theory}
\author{J\"urgen Dietel}
% \email{dietel@physik.fu-berlin.de}
\affiliation{Institut f\"ur Theoretische Physik,
Freie Universit\"at Berlin, Arnimallee 14, D-14195 Berlin, Germany}
\author{Hagen Kleinert}
% \email{kleinert@physik.fu-berlin.de}
\affiliation{Institut f\"ur Theoretische Physik,
Freie Universit\"at Berlin, Arnimallee 14, D-14195 Berlin, Germany}
\date{Received \today}
\begin{abstract}
We set up a melting model for
vortex lattices in high-temperature superconductors
based on the continuum elasticity theory.
The model is Gaussian and includes
defect fluctuations by means of a discrete-valued vortex gauge field.
We derive
the melting temperature of the lattice
and predict the size of the
Lindemann number.
Our result agrees well with  experiments for
$ {\rm YBa}_2 {\rm Cu}_3 {\rm O}_{7-\delta} $,
and with modifications also for
$ {\rm Bi}_2 {\rm Sr}_2 {\rm Ca}
{\rm Cu}_2 {\rm O}_8 $ .
We calculate the
jumps in the entropy and the magnetic induction
at the melting transition.

\end{abstract}

\maketitle

\section{Introduction}
The magnetic flux lattices
in high-temperature superconductors
can undergo a
melting transition
as was first suggested
by Nelson in 1988
 \cite{Nelson1,Nelson2}. Previously,  Brezin {\it et al.} \cite{Brezin1}
had calculated a first-order
liquid to solid phase transition by
renormalization group methods \cite{Brezin1}.
Since then detailed properties of this
transition have emerged
from various theoretical
and experimental papers.

Most prominent are computer simulations
of the Langevin equation \cite{Otterlo1}
for the dynamics of the vortices,
 or Monte Carlo simulations of XY-type model \cite{Li1} coupled to an
external magnetic field.
Analytic approaches are based mainly on the Ginzburg-Landau model
\cite{Hikami1}, or on elastic
models of the vortex lattice.
The simplest estimates
for the transition  temperature in the vortex lattice
came from an
adaption of
the famous Lindemann criterion of three-dimensional ordinary crystals
\cite{GFCM2}.
In the formulation of
 Houghton et al. and Brandt \cite{Houghton1},
the criterium states
that the vortex lattice
undergoes a melting transition
once the mean thermal displacement
$ \langle u^2\rangle^{1/2}$ reaches a certain fraction of
the lattice spacing
$ a \approx (\Phi_0/B)^{1/2} $,
 where $ \Phi_0 $ is the flux quantum and
$ B $ the magnetic
induction. The ratio
$ c_L\equiv  \langle u^2\rangle^{1/2}/a $
is the
characteristic
Lindemann number,
 which
should be independent of $B$.
Its value
is {\em not\/} predicted by Lindemann's criterion.
It must be extracted
from experiments, and is usually found to lie in the range
$ c_L \approx 0.1-0.3$.
\comment{It has been shown in \cite{Houghton1} that
$ c_L $ is, in fact, roughly universal, with only a
weak dependence on the induction of
the system.}%

The most prominent examples of high-temperature
superconductors which exhibit
vortex lattice melting are the
anisotropic compound
$ {\rm YBa}_2 {\rm Cu}_3 {\rm O}_{7-\delta} $ (YBCO),
 and the strongly layered
compound $ {\rm Bi}_2 {\rm Sr}_2 {\rm Ca}
{\rm Cu}_2 {\rm O}_8 $ (BSCCO).
Decoration experiments on BSCCO \cite{Kim1} show
the formation
of a triangular
vortex lattice, neutron scattering on
YBCO of a tilted square lattice
of vortices
\cite{Keimer1} close to the melting region, the latter being
favored by the $d$-wave symmetry of the
order parameter and the anisotropy of the crystal \cite{Won1}.
An explicit calculation of
the Lindemann number
$c_L= \langle u^2\rangle^{1/2}/a $ for YBCO can be found
in Ref. \onlinecite{Houghton1} and for BSCCO in Ref. \onlinecite{Blatter2}.

In this paper, we present a theory
which is capable of specifying
the size of the
Lindemann number  $c_L$, and
predicting corrections to the criterium.
Our theroy is based
on a simple Gaussian model
which takes into account
both lattice elasticity and defect degrees of freedom
in the simplest possible way.
The relevance of defect fluctuations for the
understanding of
melting transitions
is well-known.
For ordinary crystals,
this is textbook material
\cite{GFCM2}.
In the context of vortex melting
it was emphasized in Ref.~\cite{Marchetti1}.
For ordinary crystals,
the size of the Lindemann number has been calculated
successfully
by means of Gaussian lattice models
with elastic and defect fluctuations, which
clearly display first-order melting transitions in three dimensions
and both first-order or sequence of continuous transitions in
two dimensions.

An important quality
of
these models
 is that in the first-order case, where fluctuations are small,
they lead to a simple universal melting formula
determing the melting point in terms of the elastic constants.
The universal result is found from a lowest-order approximation,
in which one identifies the melting point with
the intersection
of the high-temperature expansion
 of the free
energy density,
dominated by defect fluctuations
with the low-temperature expansion dominated by elastic fluctuations.
The resulting
universal formula for the melting temperature
determines also the size of the Lindemann number.
Recently, the results of Ref. \onlinecite{GFCM2} for square crystals were
successfully extended to
face-centered and body-centered cubic lattices in three dimenisons
 \cite{Kleinert03}
and also to two-dimensional triangular
lattices \cite{Dietel1}.
A similar intersection criterium was also used
for the melting point
of vortex lattices
in the Abrikosov approximation of the Ginzburg-Landau model
\cite{Hikami1} useful for YBCO.
Here we shall apply our
 defect model  to
calculate the melting curves, the entropy jumps, and magnetic induction
jumps of the vortex lattices in YBCO and BSCCO.
We do not discuss in this work effects on the vortex lattice
from other sources than defect fluctuations
which can give rise to tricritical points
and glass transitions \cite{Bouquet1, Avraham1}.
Theoretical work on this subject can be found in
Refs. \onlinecite{Fischer1,Giamarchi1, Li2}.
There is in principle no problem of adding pinning in our formalism.
For simplicities, we shall confine our discussion to the defect mechanism
of melting.

The melting criterium will be derived  in Section II.
The calculation of the melting temperatures,
the entropy jumps, and the jumps of magnetic induction
for YBCO and BSCCO is carried
out in Section III.
\section{Melting Criterium}

Due to the large penetration depth $ \lambda_{ab} $ in the layers
in comparison to $ a $ we have to take into account the full non-local
elasticity constants when integrating over the Fourier space, as emphasized by
Brandt in Ref.~\onlinecite{Brandt2}.
For our Gaussian model, the partition function of the vortex lattice
can be split into
$ Z = Z_0 Z_{\rm fl} $ where $ Z_0 $ is the partition function
of the rigid lattice and $ Z_{\rm fl} $ is thermally fluctuating part
calculated via the elastic Hamiltonian plus defects.
Due to the translational invariance
of the vortex lattice in the direction of the vortices,
which we shall take to be the $z$-axis,
we may  simply
extend the
models on  square \cite{GFCM2} and triangular
 lattices  \cite{Dietel1} by a third dimension along the $z$-axis,
which we artificially discretize
to have a lattice spacing $a_3$, whose value will be fixed later.
The elastic energy is
\begin{align}
&  E_{\rm el}  =  \frac{v}{2} \sum_{{\bf x}}
 (\nabla_i u_i) (c_{11}-2c_{66}) (\nabla_i u_i)
  \label{5} \\
&   +\frac{1}{2}( \nabla_i u_j + \nabla_j u_i)\,  c_{66} \,
(\nabla_i u_j + \nabla_j u_i) +
  (\nabla_3 u_i) \, c_{44} \, (\nabla_3 u_i) .
\nonumber
\end{align}
The subscripts $ i,j $ have 
values $ 1,2$  and $l, m,n$ have values $  1,\ldots,3$. The
vectors $ u_i({\bf x})  $ are given by  the transverse displacements
of the line elements of the vortex lines with coordinate
$ {\bf x} $.
We have suppressed the spatial arguments
of the elasticity parameters, which are really
functional matrices  $
 c_{ij}({\bf x},{\bf x}')\equiv
 c_{ij}({\bf x}-{\bf x}') $.
Their precise forms were calculated by Brandt \cite{Brandt2}.
The volume
of the fundamental cell $ v $ is equal to
  $ a^2 a_3 $ (square) or $ a^2 a_3 \sqrt{3}/2 $
(triangular).
For a square lattice, the lattice derivate $ \nabla_i $
in (\ref{5}) are given by
$ \nabla_i f({\bf x})= [f({\bf x}+a {\bf e}_i) - f({\bf x})]/a $ and
$ \nabla_3 f({\bf x})= [f({\bf x}+a_3 {\bf e}_3) - f({\bf x})]/a_3 $.
For a triangular  lattice, the $xy$-part of the
lattice  has the link vectors $ \pm a \, {\bf e}_{(m)} $ with
${\bf e}_{(1,3)}=(\cos 2\pi/6,
\pm \sin 2\pi/6,0 )$ and $ {\bf e}_{(2)}=(-1,0,0) $.
The lattice derivatives around a plaquette
are defined by
$ \nabla _{(1)}f({\bf x})  =
\left[f({\bf x}+a{\bf e}_{(1)})-f({\bf x})\right]/a $,
$ \nabla _{(2)}f({\bf x}) =
\left[f({\bf x})-f({\bf x}-a {\bf e}_{(2)})\right]/a $,
$ \nabla _{(3)}f({\bf x}) =
\left[f({\bf x}-a {\bf e}_{(2)})-f({\bf x}+a {\bf e}_{(1)})\right]/a $.
From these we define
discrete cartesian lattice derivatives used in the Hamiltonian  (\ref{5})
$ \nabla_i f({\bf x})=(2/3) e_{(l)i}  \nabla_{(l)} f({\bf x}) $ and
$ \nabla_3 f({\bf x})= [f({\bf x}+a_3 {\bf e}_3) - f({\bf x})]/a_3 $
transforming like
the continuum derivative with respect to the symmetry group
 of the lattice \cite{Dietel1}. Therefore,
the  Hamiltonian (\ref{5}) has the full symmetry of the
triangular lattice and the correct
continuum elastic energy for zero lattice spacing.

The quadratic approximation to the energy (\ref{5}) is so 
far only appropriate for the  
the low-temperature classical thermodynamic behaviour. It is possible 
to extend the Hamiltonian at the {\it quadratic level}  in such a way that  
the range of applicability extends beyond the melting transition. This 
is possible by the introduction of integer-valued defect gauge fields.   
We observe that the displacement fields in (\ref{5}) 
are restricted to values within the fundamental cell.
The defect gauge fields enter to characterize the jumps of the displacement 
fields across the 
{\it Volterra surface} \cite{GFCM2,Dietel1}. As usual for gauge fields 
we choose  a minimal coupling to the lattice 
displacements.  
On account of the three lattice derivates per fundamental cell and further 
two dimensions for the displacements 
there are six independent integer-valued gauge fields per fundamental cell 
corresponding to the various defect configurations. One can eliminate 
two of them (we choose  
the defect fields corresponding to jumps in the  
$ z$-direction) by relaxing the restriction of the displacement fields  to 
the fundamental cell (elimination of gauge freedom). See the  discussion 
in Ref. \onlinecite{GFCM2} for square lattices. 
A similar consideration was 
also carried out in \cite{Dietel1} 
for the two-dimensional triangular lattice where  
the elimination of the gauge degrees of freedom is more complicated due to the 
absence  of the $ z$-direction.    
Finally, we can eliminate one more integer-valued defect 
field since the elastic energy 
in (\ref{5}) depends only on the displacements $ u_i $  
via the strain tensor  $ \nabla_i u_j +\nabla_j u_i $ \cite{GFCM2}.  
In summary, only three independent integer-valued fields have to 
be included in (\ref{5}) to obtain the elastic energy of the vortex lattice 
including defects. By taking into account the above considerations one 
can then easily determine the partition function including 
defects for the square vortex lattice 
by using the considerations in Ref. \onlinecite{GFCM2} 
and for the triangular ones by using Ref. \onlinecite{Dietel1}. 
  
By using a Hubbard-Stratonovich decoupling of the 
quadratic displacement terms in (\ref{5}),  
the stress representation  \cite{GFCM2} of the partition function becomes
\begin{widetext}
\begin{align}
& Z_{\rm fl} =  \rm {det} \left[\frac{c_{66}}{4(c_{11}-c_{66})}\right]^{1/2}
\!\!\!\!\!\rm{det} \left[\frac{1}{2\pi\beta}\right]^{5/2}
 \prod_{{\bf x}}\!
 \Bigg[ \prod_{i\leq m} \int_{-\infty}^\infty d\sigma_{im}\Bigg]
 \Bigg[\prod_{m}\sum_{n_{m}({\bf x})=-\infty}^{\infty} \Bigg]
 \Bigg[\int_{-\infty}^\infty\frac{d {\bf u}}{a}
 \Bigg]
 \exp\Bigg\{-\sum_{{\bf x}}\frac{1}{2 \beta}  \nonumber \\
 &  \times \Bigg[\sum_{i<j}\sigma _{ij}^2 +\frac{1}{2}\sum_i \sigma _{ii}^2-
\Big(\!\sum_i \sigma _{ii}\! \Big)
 \frac{c_{11}-2c_{66}}{4(c_{11}-c_{66})}\Big(\!\sum_i \sigma _{ii}\! \Big)
+\sum_i
\sigma_{i3}
\frac{c_{66}}{c_{44}}\sigma_{i3}
 \Bigg]\Bigg\}
 e^{ 2 \pi i \sum_{{\bf x}}
 (
 \sum_{i \le m}
  \nabla_{m}  u_i \sigma_{im}+  \sum_{i \le j}
  D_{ij}  \sigma_{ij}) } .
 \label{10}
 \end{align}
\end{widetext}
The parameter $ \beta $ is given by  $ \beta=
v \,c_{66}/k_B T (2\pi)^2  $. $ \sigma_{ij} $
represent stress fields \cite{GFCM2}.
The
matrix $ D_{ij}({\bf x}) $ in Eq.~(\ref{10})
is a discrete-valued local defect matrix composed of integer-valued
defect gauge fields $n_1,n_2,n_3$ for square \cite{GFCM2}
and triangular vortex lattices \cite{Dietel1}
as follows:
\begin{eqnarray}
D^{\square}_{ij} \!\!&=& \!\!\left(
 \begin{array} {cc}
n_1 & n_3 \\
n_3   & n_2
\end{array}\right),   \\
D^{\triangle}_{ij}\!\! &=&\!\!\left(
 \begin{array} {cc}
\frac{1}{2} n_1 & \frac{1}{\sqrt{3}} (n_1\!-\!n_2) \! +\!
\frac{2}{\sqrt{3}} n_3  \\
 \frac{1}{\sqrt{3}} (n_1\!-\!n_2)  \!+  \!
\frac{2}{\sqrt{3}} n_3   & -\frac{1}{2} n_1 -n_2
\end{array}      \right)  \!.~~~
\label{20}
\end{eqnarray}
The vortex gauge fields specify
the {\em Volterra surfaces\/} in units of the Burgers vectors.
By summing over all $n_{1,2,3}({\bf x})$,
the partition function $ Z_{\rm fl} $
includes all defect fluctuations,
dislocations as well as disclinations.
There is a constraint
for a vortex lattice which does not exist for ordinary three-dimensional
lattices.
Dislocations in the vortex lattice can be both screw or edge type, but
in either case the defect lines are
confined in the plane spanned by their Burger's vector
and the magnetic field \cite{Labusch1,Marchetti1}.
The reason is that the flux lines in a vortex lattice cannot be broken.
This results in the
constraint $ D_{11}= -D_{22} $ on the defect fields.

We now calculate the low-temperature expansion
of the partition function $ Z_{\rm fl} $ to lowest order, which includes only
the $ n_m=0 $-term.
By carrying out the integration
over the displacement fields
 $ u_i({\bf x}) $
in (\ref{10})
 we obtain,
as in \cite{GFCM2,Dietel1},
  the leading term in the low-temperature
expansion of the free energy
\begin{equation}
Z^{T \to 0}_{\rm fl}= \left(\frac{a_3}{a} \right)^{2N}
\frac{1}{{\rm det} \left[(2 \pi \beta){c_{44}}/{c_{66}}
\right]} \;
e^{- N \sum_{i \in\{1,6\}} l_{ii}}      ,           \label{30}
\end{equation}
where
\begin{equation}
l_{ii}=\frac{1}{2} \frac{1}{V_{\rm BZ}} \int_{\rm BZ} d^2k \, dk_3 \,
\log\left[\frac{c_{ii} a_3^2}{c_{44}}
 K^*_j K_j  + a_3^2 K^*_3 K_3     \label{40}
\right]  \,.
\end{equation}
Here $ K_m $ is the eigenvalue of $ i
 \nabla_m $.$ N $ is the number of vertices in the lattice.
The momentum
integrations in (\ref{40}) run over the Brioullin zone
of the vortex lattice
whose volume is
$ V_{\rm BZ}= (2 \pi)^3 /v $, as indicated by the subscript BZ.

Next we calculate the high-temperature expansion
$ Z^{T \to \infty}_{\rm fl} $ to lowest order. By carrying out the
integration over the displacement fields $ u_i({\bf x}) $
in (\ref{10}) and further by
summing over the defect fields $ n_m $ under the condition
$ D_{11}=-D_{22} $ mentioned above, it turns out that the stress fields
$ \sigma_{12} $ and $ \sigma_u \equiv
\sigma_{11}-\sigma_{22} $ can have only integer
numbers. Doing the integrals over the stress fields
$ \sigma_{i3} $ and  $ \sigma_g \equiv \sigma_{11}+\sigma_{22} $
we obain, in the lowest order high temperature limit, corresponding to
$ \sigma_{12}=0 $ and $ \sigma_u =0 $:
\begin{equation}
Z^{T \to \infty}_{\rm fl}=
\left(\frac{a_3}{a} \right)^{2N} \frac{C^N}{2^N}
\frac{1}{{\rm det}\left[(2\pi \beta)^2 {c_{44}}/{c_{66}}
\right]}  \,  e^{- N h}      \label{50}
\end{equation}
with
\begin{equation}
h =\frac{1}{2} \frac{1}{V_{\rm BZ}} \int_{\rm BZ} d^2k dk_3 \,
\log\left[1+\frac{c_{11}-c_{66}}{c_{44}} \,
\frac{K^*_j K_j}{ K^*_3 K_3}     \label{60}
\right]  \,.
\end{equation}
The constant $C$ has the values $ C_{\square}=1 $ for the square
vortex lattice and $ C_{\triangle}=\sqrt{3} $ for the triangular one.

In  the low-temperature expansion representing the
solid phase, defect field
configurations $ n_m \not=0 $ correspond to dislocations
and disclinations giving finite temperature corrections to
the free energy  $ -\log(Z)/k_B T $. These corrections are exponentially
small with an exponent proportinal to $ -\beta $ \cite{GFCM2,Dietel1}.
In  contrast to this corrections to the high temperature
expansion in the fluid phase corresponding to stress configurations
 $ \sigma_{12} \not=0 $ and
$ \sigma_u \not=0 $
of integer values result in temperature corrections to the free energy
which are also exponentially
small with an exponent proportional to $-1/\beta $.
The structure of the high- and low-temperature corrections to the partition
function is extensively discussed
in Refs.~\onlinecite{GFCM2,Dietel1} for ordinary crystals, and
can be easily transferred to our case of vortex lattices.
 It was shown in Refs.
\onlinecite{GFCM2,Dietel1} for the two dimensional square and triangular
as well as the three dimensional square crystal that the
exponentially vanishing higher order corrections
to the low- and high-temperature expansion of the free energy
are negligible in the determination of the transition
temperature. This is particularly true for the three
dimensional crystal (see p.~1082 in
\cite{GFCM2})
which we take as a justification to restrict our calculation
to lowest order in this paper.

From the partition function
(\ref{10}) with no defects ($ n_m=0 $) we obtain
for the Lindemann number $ c_L=\langle u^2 \rangle^{1/2}/a $
the momentum integral
\begin{equation}
c_L^2\!= \!\frac{a_3^2 }{a^2 v}  \frac{k_B T}{V_{\rm BZ}}
\int_{\rm BZ} \! \!d^2k dk_3 \frac{1}{c_{44}}\sum_{i=1,6}  \!
\frac{1}{\frac{c_{ii} a_3^2}{c_{44}}
 K^*_j K_j  + a_3^2 K^*_3 K_3 }\!.    \label{70}
\end{equation}
This  can be simplified by
taking into account
that $ c_{11} $ is much larger than $ c_{66}, c_{44} $
in the relevant regime \cite{rem1,Brandt2}.
As announced, we find the melting temperature
from
the intersection of low-and high-temperature expansions, obtained by equating
$ Z^{T \to 0}_{\rm fl}= Z^{T \to \infty}_{\rm fl} $.
By taking into account $ {\rm det}[a_3^2 \nabla^*_3  \nabla_3] =1 $
we obtain $ h, l_{11} \ll l_{66} $, and further that the $ i = 1$-term
in (\ref{70}) is much smaller than the $ i = 6 $-term. In the following
analytic discussion (but not in the numerical plots)
we neglect these small contributions.
The temperature of melting is then given by the simple formula
\begin{equation}
\frac{k_B T }{v}
\frac{1}{{\rm det}^{1/N}[c_{66}]} \, C
= \frac{ e^{-l_{66}}}{\pi} \,,                  \label{75}
\end{equation}
where $ {\rm det}[c_{66}]$ is the determinant of the
$N\times N$ functional matrix $c_{66}$.
The elastic moduli
$ c_{44} $ and $ c_{66} $
at low reduced magnetic fields $ b\equiv  B/H_{c2} <0.25 $ can be
taken from Brandt's  paper \cite{Brandt2}
\begin{eqnarray}
 c_{66} \! &  \!= \!&  \!\! \!\frac{B \phi_0 \zeta}{(8 \pi \lambda_{ab})^2}
\,, \label{80} \\
 c_{44}\! & \! = \!&  \!\! \!
\frac{B^2}{4 \pi(1\!+ \! \lambda_{c}^2 k^2\! + \!\lambda_{ab}^2 k_3^2)}
\!+ \!
\frac{B \phi_0}{32 \pi^2 \lambda_{c}^2} \!
\ln  \! \frac{ 1 \! + \!
\frac{2 \lambda^2_{c}}{\langle u \rangle^{2}} \! + \! \lambda_{ab}^2
 k^2_3}{
 1\! + \!   \lambda^2_{c}K_{\rm BZ} \!   + \!\lambda_{ab}^2 k^2_3}
 \nonumber \\
& &  + \frac{B \phi_0}{ 32 \pi^2 \lambda_{ab}^4 k_3^2}
\ln \frac{1+ \lambda^2_{ab} k_3^2/(1+\lambda^2_{ab} K^2_{\rm BZ})}{
1+ \lambda^2_{ab} k_3^2/(1+2
 \lambda^2_{ab} /\langle u \rangle^2)}    \label{90}
\end{eqnarray}
where $ \lambda_c $ is the penetration depth in the $xy$-plane,
 $ \zeta =1 $, and $ K_{\rm BZ} $ is the boundary of the
circular Brillouin zone $ K_{\rm BZ}^2=4 \pi B/\phi_0 $.
At high fields ($b>0.5$), $ c_{66} $ is altered by a factor
$ \zeta \approx 0.71 (1-b) $, and the penetration depths in $ c_{66} $,
$c_{44} $ are replaced by $ \tilde{\lambda}^2=\lambda^2/(1-b) $, where
$ \lambda $ denotes either $ \lambda_{ab} $ or $ \lambda_c $. In addition,
 the last two terms in $ c_{44} $ are replaced by $ B \phi_0/16 \pi^2
\tilde{\lambda}_c^2 $. For YBCO we have \cite{Kamal1} $ \lambda(T)=
\lambda(0)[1-(T/T_c)]^{-1/3}$, $ \xi(T)=
\xi(0)[1-(T/T_c)]^{-1/2} $ for BSCCO  \cite{Tinkham1}
   $ \lambda(T)=\lambda(0)[1-(T/T_c)^4]^{-1/2} $,
$ \xi(T)=\xi(0)[1-(T/T_c)^4]^{1/2}/[1-(T/T_c)^2]$ .
When calculating $ c_{44} $ in (\ref{90}) we have used
a momentum cutoff in the two-vortex interaction potential
$k\leq 2/\langle u ^2\rangle^{1/2}, $
 and not the inverse
of the correlation length $  1/\xi $ as in Ref. \onlinecite{Brandt2}.
The cutoff is due to
 thermal softening \cite{GFCM2}, and becomes relevant
for
$ \langle u ^2\rangle^{1/2}/ \xi \gg 1 $, or equivalently for
$ c_L \sqrt{2 \pi} \sqrt{H_{c2}(T)/B} \gg 1$, which is fulfilled in
the melting regime of BSCCO, but not for YBCO.

It remains to determine the effective lattice spacing $a_3$
along the vortex lines.
\comment{We note first that the
melting condition (\ref{75}) diverges for $a_3 \to 0 $,
whereas the displacement average (\ref{70}) remains finite in this limit.}
An elementary defect  in the vortex lattice (arising for example
from a crossing
of two vortex strings) takes place
over a typical length scale in the $z$-direction determined
from the condition that the sum of elastic
displacement energy and the energy required to stretch
the line against the line tension is minimal.
It is the elastic energy of this smallest defect that has to be taken
into account for in our model.
The energy of  an ensemble of
 dislocations is determined
by the interplay of elastic energy of small displacements and
integer-valued defect fields.
The relevant part of the free energy is given by the discretized free energy
$ -\log(Z_0 Z_{\rm fl})\, k_B T $ (\ref{10}) in which
$ a_3$ is equal to the above length scale
in the $z$-direction.
To determine this,
we insert
 the variational ansatz for the transverse displacement field
$ u_{i}= \delta_{i,1} A_0 \exp[-2 |z|/a_3] $
into the continuum version (in z direction) of the elastic energy
(\ref{5})
and approximate
$ -\nabla^2_2 \approx \langle K^2_2 \rangle \approx K_{\rm BZ}^2/4$
in $ E_{\rm el} $ and
$ K^2\approx
 \langle K^2 \rangle \approx K_{\rm BZ}^2/2$ in the elastic constants,
where
the average $ \langle \dots \rangle $ was taken
with respect to a
circular Brillouin zone. %%%weiter
The
optimal length  scale
$a_3 $ is chosen such that $ E_{\rm el} $ is minimal for
a fixed amplitude
$ A_0  \approx a $ corresponding to a typical defect elongation.

\section{Application to YBCO and BSCCo}

In the following, we  treat first the more isotropic square vortex
crystal YBCO ($ a= \sqrt{\phi_0/B}$).
From $c_{66} $ and $ c_{44} $ for YBCO,
the optimal length
 scale is given by
$ a_3= 4 a \lambda_{ab} /\lambda_c \zeta  (1-b)^{1/2}$.
When comparing the melting
criterium of the defect model
in Eq.~(\ref{75}) with the Lindemann criterium
obtained by equating the parameter (\ref{70})
to a universal number, we obtain
identical results
when taking into account
that the integrand in (\ref{70}) and in $ l_{66} $ of Eq.~(\ref{40})
receive their main contribution
from the region $ k \approx \sqrt{\langle k^2 \rangle}
\approx K_{\rm BZ}/\sqrt{2} $. We can approximate
$  k_3 \approx 0 $ in this region,  resulting in
$ a_3^2 c_{66} /a^2 c_{44} \approx 4/\pi $.
With the same approximation in
Eqs.~(\ref{40}) and (\ref{70}),
we can perform the
integrals numerically. Then we obtain
from the melting condition (\ref{75}) precisely the Lindemann criterium
in which the Lindemann number
(\ref{70}) is predicted to be
\begin{equation}
\frac{k_B T_m }{ 4  \left[ c_{44}(\frac{K_{\rm BZ}}{\sqrt{2}},0)\,
c_{66}(\frac{K_{\rm BZ}}{\sqrt{2}} ,0)\right]^{1/2} a^3}
\approx c_L^2\approx (0.18)^2
. \label{100}
\end{equation}
Denoting the spacing between the CuO$_2$ double layers
 by $ a_s $  we obtain
for the entropy jump per double layer and vortex
\begin{equation}
\Delta S_l  \approx   k_B T_m (\partial/\partial T_m) (a_s/a_3)
\ln[Z^{T \to \infty}_{\rm fl}/Z^{T \to 0}_{\rm fl}] .\label{102}
\end{equation}
Inserting (\ref{30})
and (\ref{50}), this becomes
 \begin{equation}
 \Delta S_l  \approx
  \frac{k_B T_m a_s}{a_3}  \,
\frac{\partial}{\partial T_m}
 \ln \! \frac{k_B T_m/a^3}
  { \left[ c_{44}(   \frac{K_{\rm BZ}}{\sqrt{2}} ,0)
c_{66} ( \frac{K_{\rm BZ}}{\sqrt{2}},0)
 \right]^{\frac{1}{2}}}.             \label{105}
\end{equation}
Finally, we make use of the Clausius-Clapeyron equation
which
relates the jump
of the entropy density
 across the melting
transition
 to the jump
of the magnetic induction
by
$ \Delta S_l a_3/v a_s=-(d H_m/dT) \Delta B/4 \pi  $.
Here  $ H_m $ is the
external magnetic field on the melting line.
Combining the Clausius-Clapeyron equation
with Eqs.~(\ref{90}) and (\ref{105})
we obtain, with the abbreviation $\tau_m\equiv T_m/T_c$,
the following relations near $ T_c $:
\begin{eqnarray}
 B_m(T) & \approx  &  \! \frac{ 12 \, \zeta}{16^2 \pi^4 }
\frac{(1-T/T_c)^{4/3}\,  c_L^4 \phi_0^5}{(k_BT)^2 \lambda^2_{ab}(0)\lambda^2_c(0)}  \,,
\nonumber  \\
  \Delta S_l  &  \approx &    \! \frac{\sqrt{\zeta}\, a_s}{6 a}
 \frac{\lambda_c}{\lambda_{ab}}\! \frac{k_B }{(1-\tau_m)}
\! \approx \! \frac{2.7} {10^{3}}
\frac{a_s \, c_L^2 \phi_0^2  }{T_c(1-\tau_m)^{1/3} \lambda^2_{ab}(0)} , \nonumber    \\
  \Delta B  & \approx & \! \frac{\sqrt{\zeta} \pi}{2 a \phi_0}
 \frac{\lambda_c}{\lambda_{ab}} k_B T_m \approx \frac{2.5}{10^2} \,
\frac{(1-\tau_m)^{2/3} \, c_L^2 \phi_0}{
\lambda^2_{ab}(0)}  \,.       \label{130}
 \end{eqnarray}
These results  agree
with the general
scaling results in Ref.~\cite{Dodgson1}, with the advantage that
here the
prefactors are predicted whereas those in \cite{Dodgson1}
had to be determined by fits to experimental curves
(there is only a slight discrepancy
because we use a different temperature dependence of the penetration depth).

Next, we calculate the corresponding expressions in the case of
the more layered crystal BSCCO ($a = (2^{1/2}/3^{1/4}) \sqrt{\phi_0/B} $).
First, we have to determine the
dislocation length scale $ a_3 $ in this case. For dislocation moves  we have
$ \langle u^2 \rangle^{1/2} \sim 1/ K_{\rm BZ} $.
This means that we can neglect
 the last
two terms of $ c_{44} $ in (\ref{90}),
coming from the self-energy of the vortex line,
when determining $a_3$.
Remembering this
we obtain by a similar procedure as for  YBCO
the dislocation length  scale
$ a_3 \approx  4 a \sqrt{2} \,  \lambda_{ab}/\lambda_c \sqrt{\pi}$.
From this
we find
$ a_3^2 c_{66}/ a^2 c_{44} \ll 1 $,  resulting  in $ l_{66} \approx 0 $.
By taking into account that $ B \pi^3 \lambda_{ab}^2/ 32 \phi_0 \lesssim  1 $
on the melting line we
obtain that $ c_{44}(  k ,k_3 ) $ for $ |k_3| <  \pi/a_3 $
is dominated by the last term in (\ref{90}).
Then we obtain
\begin{align}
&  c_{44}(  k ,k_3 )  \approx
\frac{B \phi_0}{32 \pi^2 \lambda_{ab}^2 (1+\lambda_{ab}^2 K^2_{\rm BZ})}
  \, \, ~{\rm for}~  k_3 \ll \frac{1}{\lambda_{ab}}\,,
  \label{135}   \\
 & c_{44}(k,k_3 )  \approx
\frac{B \phi_0 \ln(1\!+\!2B \lambda_{ab}^2/\phi_0 c_L^2) }{
 32 \pi^2 \lambda_{ab}^4 k_3^2}    ~{\rm for}~
k_3 \gg \frac{1}{\lambda_{ab}} .  \nonumber
\end{align}
By using these values we obtain by numerical integration of
(\ref{70})
\begin{align}
 &  c_L^2  \approx   \frac{k_BT_m \cdot 0.36}{a^3
 \sqrt{c_{66}(\frac{K_{\rm BZ}}{\sqrt{2}}, 0)
  c_{44}(\frac{K_{\rm BZ}}{\sqrt{2}}, 0)}} +
 \frac{k_B T_m \, a_3 \cdot 1.60 }{a^4
 c_{44}(\frac{K_{\rm BZ}}{\sqrt{2}},1/a_3)}  \label{140} \\
 &  \approx
 \frac{k_B T_m \lambda_{ab}^2 \cdot 138}{\phi^2_0 \, a}
 \sqrt{1+\lambda_{ab}^2 K_{\rm BZ}^2} +
 \frac{k_B T_m \lambda_{ab}^2 \lambda_{c}^2\cdot  137 }{\phi^2_0 \, a^3
\log(1/c_L^2)}
 \frac{\lambda_{ab}}{\lambda_{c}} .
    \nonumber
 \end{align}
The first term comes from the integration region $ |k_3| < 1/\lambda_{ab} $,
the second from the region $ 1/\lambda_{ab} < |k_3| < \pi/a_3 $
in (\ref{70}).

From our melting criterion
(\ref{75}) and the Clausius-Clapeyron equation
(where $ d H_m/dT \approx d B_m/ dT$
due to
$ B_m(T) \gtrsim H_{c1}(T) $ \cite{Zeldov1}), we obtain for BSCCO
 \begin{eqnarray}
 B_m(T) &\approx& \frac{1}{192} \frac{1}{\sqrt{3} \pi^7}
\frac{(1-(T/T_c)^4)^2}{\lambda_{ac}^2(0)   \lambda_{c}^2(0)}
\frac{\phi_0^5}{(k_BT)^2}\,,  \nonumber \\
  \Delta S_l  &\approx&     \frac{\sqrt{\pi} a_s k_B  }{4 \sqrt{2} a}
 \frac{\lambda_c}{\lambda_{ab}} \frac{1+ 3
\tau_m^4}{1-\tau_m^4}
\approx  \frac{2.9}{10^{4}} \frac{a_s (1+3 \tau_m^4)\phi_0^2 }
{T_m \lambda^2_{ab}(0)}, \nonumber    \\
  \Delta B  &\approx&  \frac{ \pi^{3/2}}{2 \sqrt{2} \,a}
 \frac{\lambda_c}{\lambda_{ab}} k_B T_m \approx \frac{1.8}{10^3}
 \frac{ (1-\tau_m^4) \phi_0}{\lambda^2_{ab}(0)}  \,.       \label{150}
 \end{eqnarray}

\begin{figure}[t]
% \hspace*{0.1cm}
\begin{center}
\psfrag{xb}{ \hspace*{-1.3cm} {\scriptsize $ T[K]$  (BSCCO)}}
\psfrag{yb}{\hspace*{-1cm} \scriptsize $ B[{\rm Gauss}]$  (BSCCO)}
\psfrag{xy}{\hspace*{-1 cm}\scriptsize $ T[K]$  (YBCO)}
\psfrag{yy}{             \hspace*{-1.3cm}
{ \scriptsize   $   B[10^4 {\rm Gauss}]$  (YBCO) } }
\includegraphics[height=5cm,width=8cm]{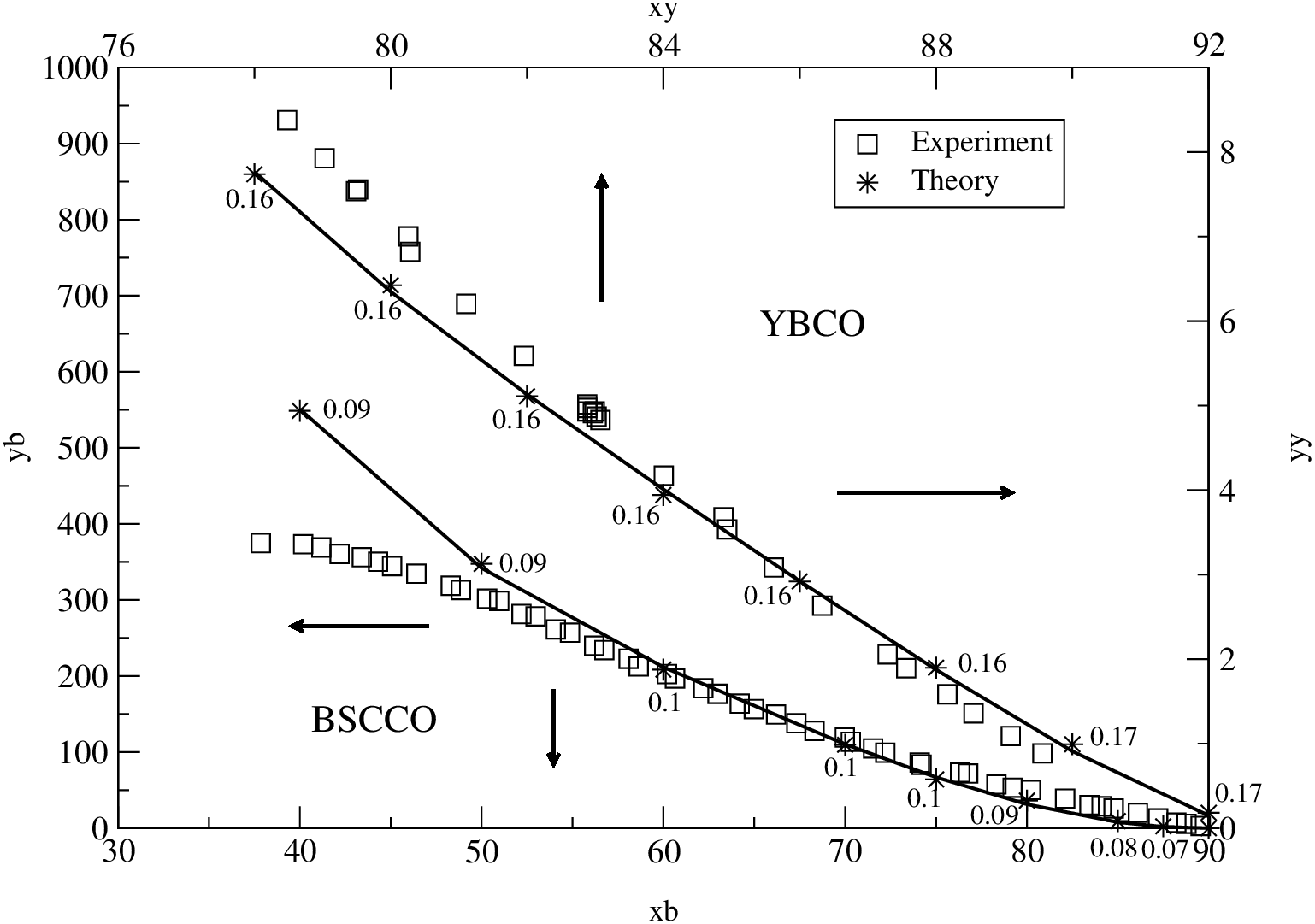}
\end{center}
 \vspace*{-0.2cm}
 \caption{Melting curve $B= B_m(T) $ for YBCO and BSCCO.
The experimental values are
 from Ref. \onlinecite{Schilling1} for YBCO and Ref. \onlinecite{Zeldov1} for
 BSCCO. The numbers on the theoretical melting curves
are the Lindemann
 numbers $ c_L $ calculated from (\ref{70}).}
\vspace*{0cm}
\end{figure}

\begin{figure}[t]
 \vspace*{0.5cm}
\begin{center}
\psfrag{T}{\scriptsize $T/T_c$}
\psfrag{deltaS}{ \hspace*{-0.4cm}
\scriptsize  $\Delta S_l [k_B] $}
\psfrag{deltaB}{\hspace*{-0.4cm}\scriptsize $\Delta B [{\rm Gauss}]$}
\includegraphics[height=6cm,width=8cm]{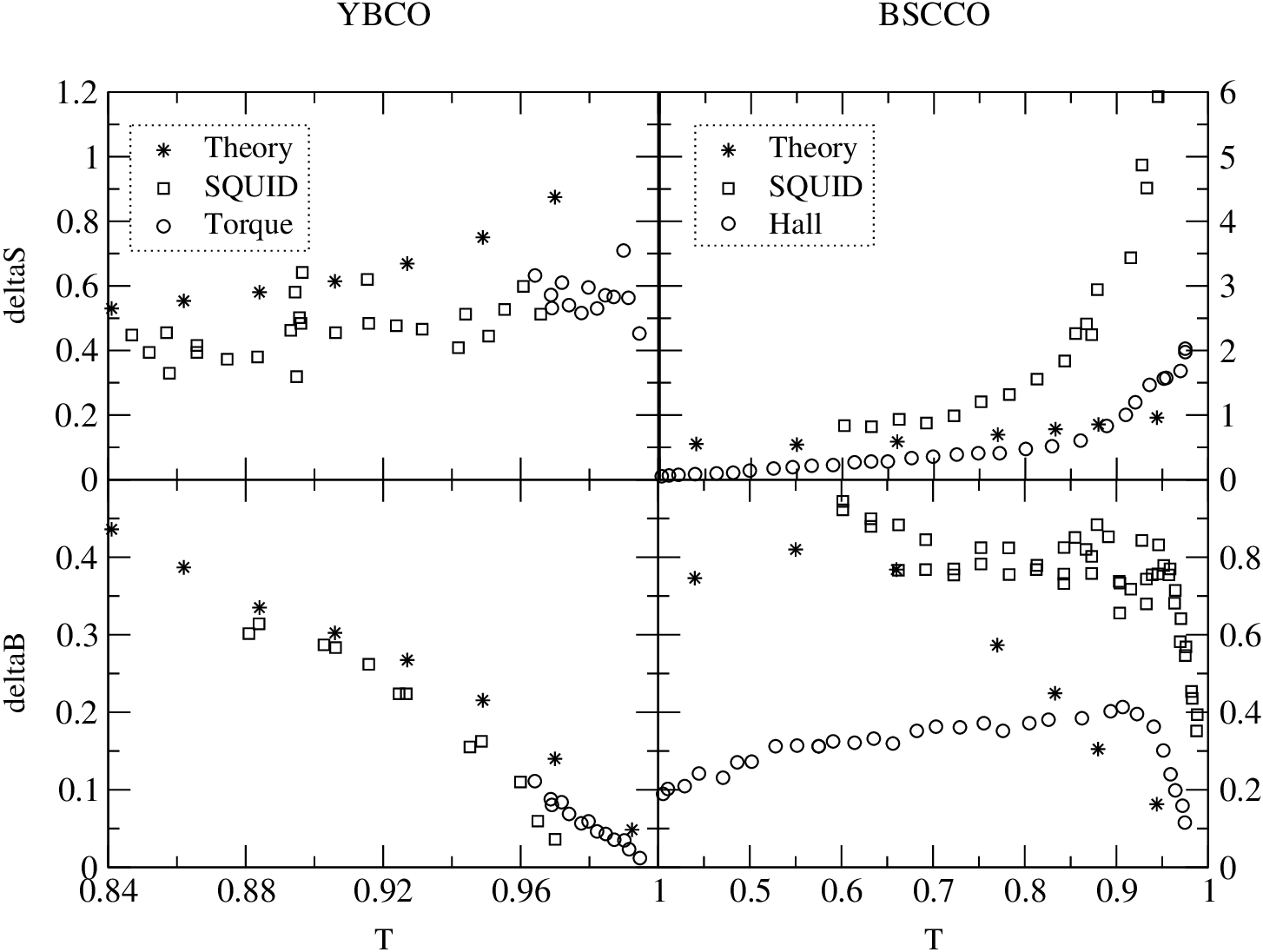}
\end{center}
\vspace*{-0.2cm}
\caption{Entropy jump per double layer per vortex $ \Delta S_l $
(first row) and jump of
magnetic induction field $ \Delta B $ (second row)
at the melting transition.
The experimental values for YBCO are from Ref. \onlinecite{Schilling1}
for $ \Delta S_l $,
Ref. \onlinecite{Welp1} for $ \Delta B $ by
SQUID
experiments (SQUID), and  Ref. \onlinecite{Willemin1} by torque measurements
(Torque).
The experimental values for BSCCO
are from Ref. \onlinecite{Kadowaki1}
by SQUID measurements (SQUID) of the magnetic field
and Ref. \onlinecite{Zeldov1} by microscopic Hall sensors (Hall).
}\vspace*{-0.4cm}
\end{figure}
Parameter values for optimal doped
YBCO (BSCCO) are given by \cite{Kamal1} $ \lambda_{ab}(0) \approx 1186 $\AA
\hspace{0.1cm}  ($ \lambda_{ab}(0) \approx 2300 $\AA), $ \xi_{ab}(0)
\approx 15 $\AA \hspace{0.1cm} ($\xi_{ab}(0) \approx 30 $\AA),
the CuO$_2$ double layer spacing $a_ s=12$\AA \hspace{0.1cm}
($ a_s=14$\AA), $ T_c=92.7 K $ ($T_c=90 K$) and
the anisotropy parameter
$ \gamma = \lambda_{c}/ \lambda_{ab} \approx 5 $ ($ \gamma \approx
200 $).

We now calculate numerically
the  melting curves, the
associated Lindemann parameter $ c_L $,
the entropy and the magnetic induction jumps
$ \Delta S_l $ and $ \Delta B $
from the intersection criterium
of the full low- and high-temperature  expressions
(\ref{30})  and (\ref{50}) without further approximation.
To accomplish this, we use the elastic constant
$ c_{11} $ given by Brandt in Ref. \cite{Brandt2}.
The intersection criterium of the low and high temperature expansion
of the partition function is then at least in the case of
BSCCO a complicated integral equation via the dependence of
$ c_{44} $ on the Lindemann parameter $ c_L $.
One can solve this integral equation by numerical methods.
The results are shown in  Fig.~1 and Fig.~2.  \\

\section{Discussion}

Our approximate analytic results
(\ref{130}) and (\ref{150}) for YBCO and BSCCO turn out to
give practically the same curves.
For comparison,
we show in both figures
the experimental curves for YBCO
of Ref.~\onlinecite{Schilling1,Welp1,Willemin1}
and for BSCCO of Ref. \onlinecite{Zeldov1,Kadowaki1}.
The good agreement
in Fig.~1
with the theoretical  curves based on  Eq. (\ref{100})
shows that the Lindemann number
is independent of the magnetic field  for YBCO
for the entire temperature range.
For BSCCO the agreement is good for smaller $B$-fields, where
the second term in (\ref{140}) indroduces some
dependence
of the Lindemann on $B$, the largest near  $B=0$.
There is some disagreement in Fig.~1
at larger $B$ and in Fig. 2
at small $B$.
This is not surprising
since
our vortex lattice
model cannot be a good approximation in these regimes.
At high $B$,
the discrepancy
comes mainly from Josephson decoupling of the layers \cite{Glazman1},
most pronounced for the strongly anisotropic
BSCCO superconductor, which leads also to large pinning effects
\cite{Blasius1}.
We think that this is also the reason for the difference
in the curves of Kadowaki {\it et al.}
in Ref. \onlinecite{Kadowaki1}
and of Zeldov {\it et al.} in Ref. \onlinecite{Zeldov1} shown in Fig.~2.
Pinning has the largest influence on the form of
the melting curve
at
 high $B$
 \cite{Khaykovich1}, resulting in a decrease of
$ \Delta S_l $ and $ \Delta B $ \cite{Li2} in the limit of
low temperatures shown by the curves
of Zeldov {\it et al.}.
Near $ B=0 $, our model does not include the
increase of the entropy coming from the thermal creation of vortices,
in addition to
the ones caused by the external magnetic field which
form the lattice \cite{Ryu1}.
Moreover, in  YBCO
order parameter fluctuations become important \cite{Ginzburg1}
which are ignored here.

Summarizing,
we have obtained
the
 melting curve, the entropy,
and the magnetic jump
from a simple lattice  defect model, and {\em derived\/} the
Lindemann rule, including the {\em size\/} of the Lindemann number, and
corrections to it. The determination of jump quantities over 
the phase transition cannot be obtained by the simple Lindemann rule. 
Our curves agree well
with the experimental curves for YBCO and BSCCO
except at zero and large magnetic fields.
The simplicity of the model has allowed us to derive
all results via analytic approximations.
Our defect model is the simplest extension
of the linear elasticity theory of
vortex displacements. We have merely added
integer-values defect gauge fields
which introduce into the elasticity theory
the rich physics of other phases
of the vortex lattice
caused by defect fluctuations,
in particular
the liquid phase and the associated melting transition.

\comment{
Our curves are in reasonable
agreement
with the
experimental data
for YBCO but not for BSCCO.
The small discrepancies for YBCO can be explained by the simplicity
of our model and the roughness
of  the approximations.
The largest discrepancy
of the melting curves for BSCCO lies in the high-field regime.
Due to the large anisotropy of BSCCO, the layers decouple
there into a two dimensional lattices of pan-cake vortices \cite{Blasius1}
by Josephson decoupling
\cite{Glazman1}. This  process
is ignored in the simple
vortex lattice model set up in this paper. In the case of the entropy
jump $ \Delta S_l $ and magnetic field jump $ \Delta B $ we obtain
in the low-temperature regime a rough agreement
of our theory with the curves of Kadowaki {\it et al.}
Ref. \onlinecite{Kadowaki1}
but not with those of Zeldov {\it et al.} Ref. \onlinecite{Zeldov1}.
The discrepancy of
the two experimental
curves  may be due to pinning which has the largest influence
on the melting curve
at low temperatures \cite{Khaykovich1}. For temperatures
near $ T_c $, the experiments give
large
values of $ \Delta S_l $ and
of $ \Delta B $,
which are not reproduced by
our
model where $ \Delta S_l $ is almost constant at large
temperatures.   This has its origin in thermally
created vortex loops in addition to the magnetically ones forming the lattice.
These vortex loops screen the effective
repulsive interaction between the magnetic induced vortices resulting in
a rise of the magnetic jump $ \Delta B $
\cite{Ryu1}.}


\begin{thebibliography}{99}
% \bibitem{shoc1} W.~Shockley, In {\it L 'Etat Solide, Proceedings of
% Neuvienne-Consail de Physique }, Brussels, 1952, Ed. R. Stoops, Solvay
% Institute de Physique, Brussels, Belgium

% \comment{\bibitem{GFCM1} H. Kleinert,{\it Gauge Fields in Condensed Matter}, % Vol. I
% {\it Superflow and Vortex lines: Disorder Fields, Phase Transition}, World
% Scientific, Singapore, 1989}

\bibitem{Nelson1}
D.~R.~Nelson, Phys. Rev. Lett. {\bf 60}, 1973 (1988).

\bibitem{Nelson2}
 D.~R.~Nelson and H.~S.~Seung, Phys. Rev. B {\bf 39}, 9153 (1989).

 \bibitem{Brezin1}
E.~Brezin, D.~R.~Nelson, and A.~Thiaville, Phys. Rev. B  {\bf 31},
7124 (1985).

\bibitem{Otterlo1}
A.~van Otterlo, R.~T.~Scalettar, and G.~T.~Zimanyi,
Phys. Rev. Lett. {\bf 81}, 1497 (1998).

\bibitem{Li1}
Y.~H.~Li and S.~Teitel, Phys. Rev. Lett. {\bf 66}, 3301 (1991);
R.~E.~Hetzel, A.~Sudb{\o}, and D.~A.~Huse,  Phys. Rev. Lett. {\bf 69}, 518 
(1992).


\bibitem{Hikami1}
S.~Hikami, A.~Fujita, and A.~I.~Larkin, Phys. Rev. B {\bf 44},
10400 (1991); D.~Li and B.~Rosenstein, Phys. Rev. Lett. {\bf 86}, 3618 (2001).

\bibitem{GFCM2} H.~Kleinert, {\it Gauge Fields in Condensed Matter}, Vol. II
 {\it Stresses and Defects}, World  Scientific, Singapore, 1989.
  (readable online at {\tt www.physik.fu-berlin.de/\~{}kleinert/re.html\#b2})

% \bibitem{Brezin} E.~Brezin {\it et al.}, Phys. Rev. B {\bf 31}, 7124 (1985).

\bibitem{Houghton1}
A.~Houghton, R.~A.~Pelcovits and A.~Sudb{\o}, Phys. Rev. B {\bf 40}, 6763 (1989); E.~H.~Brandt,
Phys. Rev. Lett. {\bf 63}, 1106 (1989).

% \bibitem{Carruzzo}
% H.~M.~Carruzzo and C.~C.~Yu, Phys. Rev. B {\bf 61}, 1521 (2000).

% \bibitem{Fisher1}
% D.~Fisher {\it et al.}, Phys. Rev. B {\bf 43}, 130 (1991).

\bibitem{Kim1}
P.~Kim , Z.~Yao, and C.~M.~Lieber, Phys. Rev. Lett. {\bf 77}, 5118 (1996).

\bibitem{Keimer1}
B.~Keimer , W.~Y.~Shih, R.~W.~Erwin, J.~W.~Lynn, F.~Dogan, and I.~A.~Aksay,
Phys. Rev. Lett. {\bf 73}, 3459 (1994);
S.~P.~Brown, D.~Charalambous, E.~C.~Jones, E.~M.~Forgan, P.~G.~Kealey,
A.~Erb, and J.~Kohlbrecher, Phys. Rev. Lett. {\bf 92}, 067004 (2004).

\bibitem{Won1}
H.~Won and K.~Maki, Phys. Rev. B {\bf 53}, 5927 (1996).

\bibitem{Blatter2}
G.~Blatter, V.~Geshkenbein, A.~Larkin, and H.~Nordborg, Phys. Rev. B {\bf 54},
72 (1996).

\bibitem{Marchetti1}
M.~C.~Marchetti and D.~R.~Nelson, Phys. Rev. B {\bf 41}, 1910 (1990);
J.~Kierfeld and V.~Vinokur,  Phys. Rev. B {\bf 61}, R14928 (2000);
J.~Lidmar, Phys. Rev. Lett {\bf 91}, 097001 (2003).

% \bibitem{GFCM2} H.~Kleinert, {\it Gauge Fields in Condensed Matter}, Vol. II
% {\it Stresses and Defects: Differential Geometry, Crystal Melting}, World
% Scientific, Singapore, 1989
% (readable online at {\tt www.physik.fu-berlin.de/\~{}kleinert/re.html\#b2}).

\bibitem{Kleinert03} H.~Kleinert and Y.~Jiang, Phys. Lett. A {\bf 313},
152 (2003).

\bibitem{Dietel1}
J.~Dietel and H.~Kleinert, Phys. Rev. B {\bf 73}, 024113 (2006).

\bibitem{Bouquet1}
F.~Bouquet, C.~Marcenat, E.~Steep, R.~Calemczuk,
W.~K.~Kwok, U.~Welp, G.~W.~Crabtree, R.~A.~Fisher, N.~E.~Phillips,
A.~Schilling, Nature (London) {\bf 411}, 448 (2001).

\bibitem{Avraham1}
N.~Avraham, B.~Khaykovich, Y,~Myasoedov,
M.~Rappaport, H.~Shtrikman, D.~E.~Feldman,
T.~Tamegai, P.~H.~Kes, M.~Li, M.~Konczykowski, K.~van der Beek , E.~Zeldov,
Nature (London) {\bf 411}, 451 (2001);
C.~J.~van der Beek, S.~Colson,
M.~V.~Indenbom, and M.~Konczykowski, Phys. Rev. Lett. {\bf 84}, 4196 (2000);
H.~Beidenkopf, N.~Avraham, Y.~Myasoedov, H.~Shtrikman,
 E.~Zeldov, B.~Rosenstein, E.~H.~Brandt and T.~Tamegai,
Phys. Rev. Lett. {\bf 95}, 257004 (2005).

\bibitem{Fischer1}
 D.~S.~Fisher, M.~P.~A.~Fisher, and D.~A.~Huse, Phys. Rev. B {\bf 43}, 130
 (1991).

\bibitem{Giamarchi1}
T.~Giamarchi and P.~LeDoussal, Phys. Rev. B {\bf 55}, 6577 (1997);
G.~P.~Mikitik and E.~H.~Brandt, Phys. Rev. B {\bf 64}, 184514 (2001);
G.~I.~Menon, Phys. Rev. B {\bf 65}, 104527 (2001).
Y.~Radzyner, A.~Shaulov,Y.~Yeshurun,  Phys. Rev. B {\bf 65}, 100513(R) (2002);
G.~P.~Mikitik and E.~H.~Brandt, Phys. Rev. B {\bf 68}, 054509 (2003);
J.~Kierfeld and V.~Vinokur, Phys. Rev. B {\bf 69}, 024501 (2004).

\bibitem{Li2}
D.~Li and B.~Rosenstein, Phys. Rev. Lett. {\bf 90}, 167004 (2003).

% \bibitem{klein ert118} H.~Kleinert, In {\it Progress in Gauge Field Theory}, e% d. by G. 't. Hooft et al., Plenum Press 1984, pp 373-401.
% ({\tt www.physik.fu-berlin.de/\~{}kleinert/re2.html\#118}).

% \bibitem{Blatter1}
% G. Blatter {\it et al.}, Rev. Mod. Phys. {\bf 66}, 1125 (1994).

\bibitem{Brandt2}
 E.~H.~Brandt, Rep. Prog. Phys. {\bf 58}, 1465 (1995); see also
G. Blatter, M.~V.~Feigel'man, V.~B.~Geshkenbein, A.~I.~Larkin,
and V.~M.~Vinokur, Rev. Mod. Phys. {\bf 66}, 1125 (1994).


% \bibitem{Houghton1}
% A.~Houghton, R.~A.Pelcovits, and A.~Sudbo, Phys. Rev. B {\bf 40}, 6763
% (1989).

% \bibitem{Carruzzo1}
% H.~M.~Carruzzo and C.~C.~Yu, Phys. Rev. B {\bf 61}, 1521 (1528).

% \bibitem{Marchetti2}
% M.~C.~Marchetti and L.~R.~Radzihovsky, Phys. Rev. B {\bf 59}, 12001 (1999).

\bibitem{Labusch1}
R.~Labusch, Physics Letters {\bf 22}, 9 (1966).

\bibitem{rem1}
By using the elastic constants from  Ref. \onlinecite{Brandt2} we obtain that
$ c_{11} \gg c_{44}, c_{66} $ almost everywhere on the melting line except
in a small vicinity near $ T \approx T_c $.
Here we take into account that $ |k_3| < \pi/a_3 \approx
\pi  \lambda_c/\lambda_{ab} a  $ and further that
$ B \lambda_{ab}^2 /\phi_0 \gg 1/16 \pi $ is fulfilled almost everywhere
on the melting line (except in a region where
$ |T/T_c-1|\lesssim 0.02 $ for BSCCO, the corresponding
region for YBCO is even smaller).


\bibitem{Kamal1}
 S.~Kamal, D.~A.~Bonn, N.~Goldenfeld, P.~J.~Hirschfeld, R.~Liang,
and W.~N.~Hardy, Phys. Rev. Lett. {\bf 73}, 1845 (1994).

\bibitem{Tinkham1}
M.~Tinkham, {\it Introduction to Superconductivity}, McGraw-Hill, New York,
1996.

% \bibitem{Brandt2}
% E.~H.~Brandt, J. Low Temp. Phys. {\bf 26}, 709 (1977).

% \bibitem{Brandt3}
% E.~H.~Brandt, J. Low Temp. Phys. {\bf 26}, 735 (1977).

% \bibitem{Mikitik1}
% G.~P.~Mikitik and E.~H.~Brandt, Phys. Rev. B {\bf 68}, 054509 (2003).

\bibitem{Dodgson1}
M.~J.~W.~Dodgson, V.~B.~Geshkenbein, H.~Nordborg, and G.~Blatter, Phys. Rev. Lett {\bf 80}, 837 (1998).

\bibitem{Zeldov1}
E.~Zeldov,  D.~Majer, M.~Konczykowski, V.~B.~Geshkenbein, V.~M.~Vinokur,
and H.~Shtrikman, Nature (London) {\bf 375}, 791 (1995).

\bibitem{Schilling1}
A.~Schilling, R.~A.~Fisher, and G.~W.~Crabtree,
Nature (London) {\bf 382}, 791 (1996).

\bibitem{Welp1}
U.~Welp, J.~A.~Fendrich, W.~K.~Kwok, G.~W.~Crabtree, and B.~W.~Veal,
Phys. Rev. Lett. {\bf 76}, 4809 (1996).

\bibitem{Willemin1}
M.~Willemin, A.~Schilling, H.~Keller, C.~Rossel, J.~Hofer, U.~Welp,
W.~K.~Kwok, R.~J.~Olsson, and G.~W.~Crabtree,  Phys. Rev. Lett. {\bf 81}, 4236 (1998).

\bibitem{Kadowaki1}
K.~Kadowaki and K.~Kimura, Phys. Rev. B {\bf 57}, 11674 (1998).

\bibitem{Glazman1}
L.~I.~Glazman and A.~E.~Koshelev, Phys. Rev. B {\bf 43}, 2835 (1991).

\bibitem{Blasius1}
T. Blasius, Ch.~Niedermayer, J.~L.~Tallon, D.~M.~Pooke, A.~Golnik,
and C.~Bernhard, Phys. Rev. Lett. {\bf 82}, 4926 (1999).

\bibitem{Khaykovich1}
B.~Khaykovich, M.~Konczykowski, E.~Zeldov,R.~A.~Doyle,
D.~Majer, P.~H.~Kes and T.~W.~Li, Phys. Rev. B {\bf 56}, R517 (1997).

\bibitem{Ryu1}
S.~Ryu and D.~Stroud, Phys. Rev. B {\bf 57}, 14476 (1998).

\bibitem{Ginzburg1}
V.~L.~Ginzburg, Fiz. Twerd. Tela {\bf 2}, 2031 (1960)
[Sov. Phys. Solide State {\bf 2}, 1824 (1961)].

\end{thebibliography}
\end{document}